\documentclass[prd,a4paper,showpacs,byrevtex]{revtex4}
\usepackage{graphicx}
\usepackage{dcolumn}
\usepackage{amsmath}
\usepackage{array}
\usepackage{bm}
\usepackage{amssymb}
\usepackage{amsfonts}
\usepackage[noautoscale,vcentermath]{youngtab}

\begin{document}
\title{Glueball phenomenology and the relativistic flux tube model}
\author{Fabien \surname{Buisseret}}
\thanks{F.R.S.-FNRS Research Fellow}
\email[E-mail: ]{fabien.buisseret@umh.ac.be}
\author{Vincent \surname{Mathieu}}
\thanks{IISN Scientific Research Worker}
\email[E-mail: ]{vincent.mathieu@umh.ac.be}
\author{Claude \surname{Semay}}
\thanks{F.R.S.-FNRS Research Associate}
\email[E-mail: ]{claude.semay@umh.ac.be}
\affiliation{Groupe de Physique Nucl\'{e}aire Th\'{e}orique,
Universit\'{e} de Mons-Hainaut,
Acad\'{e}mie universitaire Wallonie-Bruxelles,
Place du Parc 20, BE-7000 Mons, Belgium}

\date{\today}
\begin{abstract}
The relativistic flux tube model is an effective description of confined quarks and gluons in which the confining interaction is carried by the flux tube, a Nambu-Goto string. We first show that the relativistic flux tube model can be applied to glueballs seen as bound states of transverse constituent gluons. After a comparison of that approach with usual spinless Salpeter Hamiltonians, we compute glueball masses and decay widths. Comments about the $\eta$-$\eta'$-pseudosclar glueball problem, the glueball--Pomeron conjecture, and finite-temperature effects are finally given. We also point out the existence of a duality between open- and closed-flux tube models of glueballs.
\end{abstract}
\maketitle

\section{Introduction}
Quantum chromodynamics (QCD) allows the existence of pure gauge states called glueballs. Although much effort has been devoted to find a clear experimental glueball signal, no unambiguous candidate is known so far~\cite{crede}. Nevertheless, glueballs are intensively studied within various theoretical frameworks: lattice QCD~\cite{latti0,latti1,latti2}, Coulomb gauge QCD~\cite{couglu}, AdS/QCD~\cite{adsq}, QCD sum rules~\cite{sumrul}, potential models~\cite{barnes,glupot,glupot2,kaid,gluh1}, \textit{etc}. We refer the interested reader to Ref.~\cite{review} for a review on theoretical glueball physics. Nowadays, the glueball mass spectra computed in lattice QCD have a special status since they are often taken as input to fit the parameters of the existing models. Indeed, clear experimental data related to glueballs are still missing. 

The present work is devoted to an effective approach of QCD that has poorly been applied to glueballs so far: the relativistic flux tube model~\cite{tf1,bramb95,ch_aux_bib}. Actually, in the meson case, it follows from nonperturbative QCD in the quenched approximation that the confining interaction between a quark and an antiquark pair is due at the dominant order to a Nambu-Goto string linking them and carrying both energy and angular momentum~\cite{bramb95}. This agrees with lattice QCD simulations showing that the chromoelectric field between a static quark-antiquark pair is roughly constant on a straight line joining these two particles (see for example the early study \cite{balitf}, and the more recent works \cite{ichie,Bissey}). Calculations within the dual Ginzburg-Landau theory also support the existence of a flux tube~\cite{Koma}. The relativistic flux tube model is not a potential approach since the interaction is mediated by a dynamical object. In particular, not only light and heavy mesons can be described in that framework~\cite{sema95,expe}, but also hybrid mesons when the flux tube is in an excited state~\cite{alle}. Here we extend the relativistic flux tube model to glueballs, assuming that these exotic hadrons can be described as bound states of transverse constituent gluons. We point out that using longitudinal constituent gluons leads to spurious states with respect to those observed in lattice QCD, and to a mass spectrum that cannot agree with the lattice data in the $C=-$ sector~\cite{glupot2,expe}.

After having recalled how to build quantum states of transverse gluons in section~\ref{helfor}, we present the relativistic flux tube model and its generalization to gluonic bound states in section~\ref{ftm}. We also study the Regge trajectories, that are a consequence of the relativistic flux tube model in the ultrarelativistic limit, in section~\ref{regge}. Then we compute the glueball mass spectrum in section~\ref{gluspec} and estimate the glueball decay widths in section~\ref{decay}. Finally, we discuss about the gluonic content of the $\eta$-$\eta'$ resonances, the glueball-Pomeron conjecture, and finite-temperature effects in sections~\ref{etaetap}, \ref{glupom}, and \ref{finiT}. Our results are summarized in section~\ref{Conclu}.

\section{Bound states of transverse gluons}\label{helfor}

\subsection{Generalities}
The main goal of this work is to build an effective model of glueball based on the relativistic flux model in which the constituent gluons are transverse, that is with helicity-1. As we argued in~\cite{gluh1,glub}, the helicity formalism of Jacob and Wick~\cite{jaco} can be successfully applied to describe a bound state of two such gluons. It is thus important to show here how to build glueball helicity states before using them as basis states for computations within the relativistic flux tube model.

Let us recall the main points of the helicity formalism, introduced in~\cite{jaco}. We first introduce $\left|\psi(\vec p,\lambda)\right\rangle=a^{\dagger}_{\lambda}(\vec p\ )\left|0\right\rangle$ the quantum state of a particle with momentum $\vec p$, spin $s$, and helicity $\lambda$. If the particle is transverse (with a helicity degree of freedom), only $\lambda=\pm s$ is allowed, while the $(2s+1)$ projections from $-s$ to $+s$ are allowed if the particle has a spin degree of freedom. Then it can be deduced from~\cite{jaco} that the quantum state
\begin{eqnarray}
\left|\lambda_1,\lambda_2;J,M\right\rangle&=&\left[\frac{2J+1}{4\pi}\right]^{\frac{1}{2}}\int^{2\pi}_0d\phi\int^\pi_0d\theta\, \sin\theta\ {\cal D}^{J*}_{M,\lambda_1-\lambda_2}(\phi,\theta,-\phi)\, R(\phi,\theta,-\phi)\ a^{\dagger}_{\lambda_1}(\vec p)a^{\dagger}_{\lambda_2}(-\vec p)\left|0\right\rangle
\end{eqnarray}
represents a two-particle state in the rest frame of the system which is also an eigenstate of the total spin $\vec J$, \textit{i.e.} $\vec J^{\, 2}=J(J+1)$ and $J_z=M$. In the above definition, $R(\alpha,\beta,\gamma)$ is the rotation operator of Euler angles $\{\alpha,\beta,\gamma\}$ and ${\cal D}^{J}_{M,\lambda}(\alpha,\beta,\gamma)$ are the Wigner $D$-matrices. The coordinates $\{\theta,\phi\}$ are the polar angles of $\vec p$. It is worth recalling the inequality
\begin{equation}
J\geq|\lambda_1-\lambda_2|,
\end{equation}
coming from usual spin-coupling rules. Sill by applying the formalism of~\cite{jaco}, it can be checked that the state
\begin{equation}\label{hstate}
\left|\lambda_1,\lambda_2;J^P,M,\epsilon\right\rangle=\frac{1}{\sqrt 2}\left[	 \left|\lambda_1,\lambda_2;J,M\right\rangle+\epsilon\left|-\lambda_1,-\lambda_2;J,M\right\rangle\right]
\end{equation}
is also an eigenstate of the parity operator with the eigenvalue
\begin{equation}
    P=\epsilon\, \eta_1\eta_2(-1)^{J-s_1-s_2},
\end{equation}
$\eta_i$ and $s_i$ denoting the intrinsic parity and spin of particle $i$ respectively. Let us note that only the relative sign of $\lambda_1$ and $\lambda_2$ is relevant.

Either particles with spin or helicity can be described within this formalism. When both particles have a spin degree of freedom, the helicity basis, spanned by the helicity states~(\ref{hstate}), is equivalent to a standard
$\left|^{2S+1}L_J\right\rangle$ basis up to an orthogonal transformation \cite{gie}. When at least one of the particles is
transverse, both basis are no longer equivalent but the helicity states can still be expressed
as particular linear combinations of $\left|^{2S+1}L_J\right\rangle$ states \cite{gluh1,jaco}. It is worth mentioning that only the spin-angular part of the helicity states is fixed by the above geometrical construction. The radial part of the system needs a particular dynamical model to be determined.

Although we will mainly focus on two-gluon states in the present work, it is worth mentioning that three-body helicity states can also be built following a two-by-two recoupling scheme~\cite{gie,wick3}. The idea is to write first the two-body helicity state~(\ref{hstate}) corresponding to, say, particles 2 and 3 in their rest frame. Then, the Lorentz-boosted helicity state of the (2,3) cluster can be coupled to particle 1 with the same technique. The boost is needed for the three-body helicity state to be expressed
in the three-body rest frame, not in the rest frame of the cluster. That recoupling procedure can in principle be generalized to ${\cal N}$-body systems.

\subsection{Two-gluon states}

Transverse gluons are such that $\lambda_i=\pm1$. Since $\epsilon=\pm1$, there are thus four independent helicity states of the form~(\ref{hstate}), and the Pauli principle must be satisfied. The color wave function of a two-gluon system, $[\bm 8,\bm 8]^{\bm 1}$, is symmetric and enforces a positive charge conjugation; one has thus to ask the two-gluon states to be totally symmetric. The symmetrization leads in general to selection rules on the total spin of the system~\cite{jaco}. We have shown in Ref.~\cite{gluh1} that, in the case of two gluons, the symmetrized helicity states read
\begin{align}\label{glustate}
\left|S_+;(2k)^+\right\rangle&=\left|1,1;(2k)^+,M,1\right\rangle,& \left|D_+;(2k+2)^+\right\rangle	&=\left|1,-1;(2k+2)^+,M,1\right\rangle,\\
	\left|S_-;(2k)^-\right\rangle	&=\left|1,1;(2k)^-,M,-1\right\rangle,& \left|D_-;(2k+3)^+\right\rangle	&=\left|1,-1;(2k+3)^+,M,-1\right\rangle,\nonumber
\end{align}
with $k\in\mathbb{N}$. The $S$ ($D$)-states will be referred to as helicity singlets (doublets) in the following. It appears that there are no $J^{PC}=1^{P+}$ two-gluon states, in agreement with Yang's theorem forbidding the decay of vector mesons in two photons~\cite{yang}. This fact is also consistent with the absence of low-lying $1^{P+}$ glueballs in the lattice QCD results~\cite{latti0,latti1,latti2}.

Once the glueball helicity states~(\ref{glustate}) are known, it is possible to compute matrix elements. In particular, we have checked in Ref.~\cite{gluh1} that none of the usual operators (identity, spin-orbit, tensor, \dots) induce couplings between the $S_+$ and $D_+$ states, that possess the same $J^{PC}$ quantum numbers. Moreover, the average value of the square orbital angular momentum takes the simple form
\begin{equation}\label{lsdef}
    \left\langle \vec L^{\, 2}\right\rangle=J(J+1)+2\lambda_1\lambda_2.
\end{equation}
It can be expected that $\left\langle \vec L^{\, 2}\right\rangle $ roughly sets the energy scale of a given state: The more rotational energy is contained in a glueball, the heavier the state should be. This suggests the following mass ordering~\cite{barnes}: $0^{\pm+}$, $2^{++}$, $2^{-+}$, $3^{++}$, \textit{etc.}, in agreement with the lightest glueball states observed in lattice QCD~\cite{latti0}. Two is indeed the minimal number of gluons needed to make a color singlet, and it is rather intuitive to see the lightest glueballs as two-gluon bound states.

\subsection{Many-gluon states}\label{fourglu}

The situation is more complicated for three-gluon glueballs. First of all, two color singlets are possible: One symmetric (leading to $C=-$) and one antisymmetric (leading to $C=+$). The corresponding color wave functions are $[\left[\bm 8,\bm 8\right]^{\bm 8_s},\bm 8]^{\bm 1_s}$ and $[\left[\bm 8,\bm 8\right]^{\bm 8_a},\bm 8]^{\bm 1_a}$ respectively. The low-lying $C=-$ glueballs are well-known from lattice computations. We gave in Ref.~\cite{glub} several arguments favoring a three-gluon interpretation of these states, the most obvious ones being that negative charge conjugations cannot be reached by a two-gluon bound state, and that the lightest $C=-$ glueballs are heavier than the $C=+$ ones. Moreover, the lowest-lying bound states of three gluons should be $J^C=1^-$ and $3^-$ ones, as observed in lattice QCD~\cite{glub}. Nevertheless, writing completely symmetrized three-gluon states in the helicity formalism is a complex problem that would deserve a separate paper; it is beyond the scope of the present work.

It is worth saying that building a (pseudo)scalar state from a totally symmetric three-gluon system might be problematic. This case is indeed formally identical to a three-photon system. It has been previously shown that a planar three-photon system where the photons have equal energies and momenta separated by 120$^{\rm o}$ from one another cannot be in a (pseudo)scalar state~\cite{fumi}. Such a situation is analogous to low-lying three-gluon states; no light $0^{P-}$ glueball is indeed observed. However, asymmetric three-photon configurations, that could correspond to highly excited glueballs, can be in such a state~\cite{fumi2}. Consequently, if there is a three-gluon glueball with $J^C=0^-$, it must be particularly heavy. In agreement with that picture, the heaviest-known glueball on the lattice is a $0^{+-}$ state. But we argued in Ref.~\cite{glub} that, rather than a highly excited three-gluon state, the $0^{+-}$ glueball could also be a low-lying four-gluon state with the nontrivial color wave function $[\left[\bm 8,\bm 8\right]^{\bm{10}_a},\left[\bm 8,\bm 8\right]^{\overline{\bm{10}}_a}]^{\bm 1}$, imposing the mixed symmetry $\yng(2,2)$ to the spin-space wave function. Notice that $C=+$ glueballs in the mass range of the $C=-$ ones are currently not known from lattice computations since one- and two-glueball states cannot be distinguished in this sector~\cite{latti0}. The most interesting states to study there would be the $1^{P+}$ glueballs, that cannot be made of two gluons.

\subsection{Comments on gluelumps}

Gluelumps have been proposed in lattice QCD as a first approach to model gluino-gluon bound states~\cite{foster}. They are pure gauge states, as glueballs, but this time a static, scalar, color-octet source is present in the system. It appears that the gluelump masses are particularly low (the static source is decoupled from the system), actually lower than the lightest glueballs; the lightest gluelump is indeed a $1^{+-}$ state of mass 0.87$\pm0.15$~GeV~\cite{gl1}. This can be understood in terms of constituent gluons because the presence of the color-octet source allows for only one gluon to be bound in a color singlet. Gluelumps seen as one-gluon states have been studied in detail within the Coulomb gauge QCD approach~\cite{guo}, but also with a spinless Salpeter Hamiltonian~\cite{lumpb}. The gluelump helicity states can be found in this last reference and read
\begin{eqnarray}\label{hsgl1}
| T_+;(J\geq 1)^{P}\rangle &= \left|0,1;J^P,M,1\right\rangle\qquad &\textrm{with} \qquad P=(-)^{J}, \nonumber\\
| T_-;(J\geq 1)^{P}\rangle &= \left|0,1;J^P,M,-1\right\rangle \qquad &\textrm{with} \qquad P=(-)^{J+1}.
\end{eqnarray}
They have a negative charge conjugation and are such that $ \left\langle \vec L^{\, 2}\right\rangle=J(J+1)$; formula~(\ref{lsdef}) also holds in this case. We point out that using longitudinal gluons would lead to a low-lying $0^{--}$ gluelump, in disagreement with the lattice data.

Interestingly, one-gluon states have been studied in lattice QCD more than twenty years ago in order to estimate the constituent gluon mass~\cite{bern}. The following idea has been applied: The potential between two static color-octet sources should saturate at the energy at which a pair of gluons can be created from the vacuum. The energy of a confined gluon is thus half that energy. One arrives at an estimate of 500-800~MeV~\cite{bern}. In the bag model, the same quantity can be computed and is given by $740\pm100$~MeV~\cite{dono}, a value compatible with the $1^{+-}$ gluelump. Interpreting the mass of that gluelump as a constituent gluon mass, one estimates the low-lying two- and three-gluon glueballs to be around $1.74\pm0.30$~GeV and $2.61\pm0.45$~GeV, in agreement with lattice computations~\cite{latti0,latti1}.

\section{The relativistic flux tube model}\label{ftm}

Once glueball helicity states are known, a Hamiltonian remains to be specified. To this aim, we present in this section the relativistic flux tube model. Although its original formulation, concerning mesons, relies on purely phenomenological grounds~\cite{tf1,tf2}, the relativistic flux tube model has been shown to be a well-defined approximation of the nonperturbative QCD interactions between a quark and an antiquark~\cite{bramb95,QCDstring}. We recall here the main aspects of this model, and refer the interested reader to these last two references for more details.

\subsection{Two-body case}

Let us begin by the case of a meson. The study of the quark-antiquark Green function in QCD involves the Wilson loop whose study eventually leads to crucial information about nonperturbative effects in QCD. In particular, an effective quark-antiquark Lagrangian can be obtained from the Wilson loop provided that the quenched approximation is made and that only the long-range, spin-independent, part of the interactions is considered. The spin- and short-range effects are actually not expected to be dominant with respect to the long-range interactions corresponding to the confinement. This covariant effective Lagrangian reads \cite{bramb95,QCDstring}
\begin{equation}\label{lag1}
	{\cal L}=-m_1\, \sqrt{\dot{\rm x}^2_1}-m_2\, \sqrt{\dot{\rm x}^2_2}-a\, \int^{1}_{0}d\theta\, \sqrt{(\dot{\rm w}\, {\rm w}')^2-\dot{\rm w}^2\, {\rm w}'^2},
\end{equation}
with the boundary conditions
\begin{equation}\label{bound}
	\left.{\rm w}^\mu\right|_{\theta=1}={\rm x}^\mu_1,\quad 	\left.{\rm w}^\mu\right|_{\theta=0}={\rm x}^\mu_{2}.
\end{equation}
The first two terms of Lagrangian~(\ref{lag1}) are kinetic terms corresponding to the quark and the antiquark seen as massive spinless color sources, whose position in Minkowski spacetime is given by the four-vector ${\rm x}_{i}(\tau)$. $\tau$ is a timelike evolution parameter of the system, and the dotted quantities denote a derivation with respect to this parameter.  The third term of Lagrangian~(\ref{lag1}) is a Nambu-Goto Lagrangian. It describes a string of generic energy density $a$ linking the constituent particles through the boundary conditions~(\ref{bound}). We notice that the string coordinates ${\rm w}(\theta,\tau)$ also depend on a spacelike parameter labeling its different points; the prime denotes a derivation with respect to $\theta$. This string is the so-called relativistic flux tube (or QCD string). It is a dynamical object arising from the nonperturbative part of the strong interaction.

As mentioned in the introduction, the formation of flux tubes in QCD is supported in the meson and baryon cases by lattice QCD simulations. Moreover, it is of particular interest for the present study to stress that recent lattice QCD computations strongly favor the flux-tube picture in systems with constituent gluons like quark-antiquark-gluon~\cite{Bicu1} and three-gluon systems~\cite{Bicu2}. Notice that, in all these lattice studies, quarks and gluons are seen as static color sources. Finally, Lagrangian~(\ref{lag1}) has been shown to be valid also for two-gluon systems in Ref.~\cite{kaid}. The relativistic flux tube picture thus appears to be an interesting effective approach in the case of hadronic bound states. A comment on the value of $a$ should be done at this stage. Let us denote $\sigma$ the fundamental string tension, \textit{i.e.} the energy density of a flux tube linking a quark and an antiquark. Then, several lattice studies suggest that the adjoint string tension, that is energy density of a flux tube linking two static color-octet sources, should be given by $\sigma_a=(9/4)\sigma$~\cite{Bicu1,Bicu2,cas}. In other words, we consider that the string tension scales as the quadratic color Casimir operator: This is the Casimir scaling hypothesis.

It is convenient for further calculations to work in the temporal (or instantaneous) gauge, that is ${\rm x}^0_1={\rm x}^0_2=\tau$. In that gauge, the solution of the string equations of motion is simply given by the straight line ${\rm w}=\theta\, {\rm x}_1+(1-\theta)\, {\rm x}_{2}$; see for example~\cite{Allen}. Notice that the linearity of the flux tubes is in agreement with lattice QCD~\cite{balitf,ichie}. Once Lagrangian~(\ref{lag1}) is rewritten in the temporal gauge and with a straight flux tube, the relativistic flux tube equations can be obtained in the Hamiltonian formalism~\cite{tf1,tf2,bramb95}. In their quantized version, those equations read, in the rest frame and in the equal mass case ($m_1=m_2=m$)~\cite{tf1,sema95}
\begin{eqnarray}
\label{tf_equa1}
\frac{2}{r}\left\langle\sqrt{ \vec L^{\, 2}}\right\rangle&=&\{v\gamma,W_{r}\}+a\{r,f(
v)\}, \\
\label{tf_equa2}
H&=&\{\gamma,W_{r}\}+\frac{a}{2}\left\{r,\frac{\arcsin
v}{v}\right\},
\end{eqnarray}
where $\{A, B\}=AB+BA$ denotes the anticommutator, and where
\begin{equation}
	W_{r}=\sqrt{p^{2}_{r}+m^{2}},\qquad f(v)=\frac{\arcsin v-v\sqrt{1-v^{2}}}{4v^2},\qquad \gamma=\frac{1}{\sqrt{1-v^2}}.
\end{equation}
$v$ represents the modulus of the transverse velocity of the particles, and commutes neither with the radius $r$ nor with the radial momentum $p_r$. That is why various symmetrizations have to be performed. The complexity of these last equations comes from the fact that the contribution of the flux tube does not reduce to a static potential: This object is dynamical, and carries orbital angular momentum as well as energy. In consequence, $v$ cannot be extracted from~(\ref{tf_equa1}) and the Hamiltonian cannot be written in a closed form. One can observe in~(\ref{tf_equa2}) that the linearly rising potential commonly used to describe confinement is supplemented by terms involving $v$, thus by relativistic effects due to the flux tube. Also the kinetic part is logically affected by these relativistic effects. We mention for completeness that equations generalizing~(\ref{tf_equa1}) and (\ref{tf_equa2}) can be obtained in the meson case when the spin of the quark and of the antiquark is no longer neglected~\cite{Allen2,jug}. The resulting equations are quite complicated and we will not write them here explicitly. One actually gets couplings between the quark spin and the flux tube that can be understood as a Thomas precession in the color field generated by the flux tubes~\cite{Allen2}. We point out that, for computational convenience, the following assumption will be made
\begin{equation}\label{lapp}
\left\langle \sqrt{\vec L^{\, 2}}\right\rangle\approx\sqrt{\left\langle \vec L^{\, 2}\right\rangle}.	
\end{equation}
 
Various limits of the equations~(\ref{tf_equa1}) and (\ref{tf_equa2}) can be considered. When $\left\langle \vec L^{\, 2}\right\rangle=0$, then one has trivially $v=0$ and the Hamiltonian of the problem simply becomes $H=2\sqrt{\vec p^{\, 2}+m^{2}}+ar$ since $\vec p^{\, 2}=p^2_r+\vec L^{\, 2}/r^2$, that is a widely used spinless Salpeter Hamiltonian with linear confinement. In the nonrelativistic limit, an expansion in $1/m$ can be performed, and the resulting Hamiltonian is $2m+\vec p^{\, 2}/m+ar$, a Schr\"odinger equation with linear confinement. On the contrary, if $m=0$ and if $\left\langle \vec L^{\, 2}\right\rangle$ is defined as $\ell(\ell+1)$, it can be shown by a semiclassical analysis that $H^2\approx 2\pi a\ell$ at large $\ell$~\cite{tf2}. These are the well-known Regge trajectories, observed in particular in light meson spectroscopy. Finally, if the dynamical flux tube contribution is neglected in~(\ref{tf_equa1}), one is led to the standard spinless Salpeter Hamiltonian~\cite{tf1}
\begin{equation}\label{ssh}
H_0=2\sqrt{\vec p^{\, 2}+m^{2}}+ar.	
\end{equation}
The corrective term
\begin{equation}\label{pft}
	\Delta H_{ft}=-\frac{a\, \left\langle \vec L^{\, 2}\right\rangle}{\mu\, r\, (6\mu+ar)},\qquad {\rm with}\qquad \mu=\left\langle \sqrt{\vec p^{\, 2}+m^{2}}\right\rangle,
\end{equation}
can also be added in perturbation to take into account the dynamical contribution of the flux tube at dominant order. The Hamiltonian $H_0+	\Delta H_{ft}$ defines the so-called perturbative flux tube model, that has already been studied and successfully applied to describe light mesons in~\cite{expe,bada}.

For two-gluon glueballs, one has obviously $m=m_g$ with $m_g$ the gluon mass. Moreover, we set $a=\sigma_a=(9/4)\sigma$ and $\left\langle \vec L^{\, 2}\right\rangle$ is given by~(\ref{lsdef}).

\subsection{Many-body generalization}

The relativistic flux tube model can be generalized to many-body systems as follows. From the Casimir scaling hypothesis, one considers that each color source generates a straight flux tube whose energy density is proportional to its quadratic color Casimir operator. Then, the flux tubes have to meet in one or several points such that the total energy contained in those flux tubes is minimal. In baryons for example, the junction point is called the Toricelli (or Fermat or Steiner) point of the triangle made by the quarks, leading to a Y-junction in agreement with lattice computations~\cite{ichie,taka1,taka2}. For multiquark states (tetraquarks, pentaquarks,\dots), lattice studies indicate that the confining potential is compatible with a multi Y-junction flux-tube configuration~\cite{oki1,oki2}, again minimizing the total flux-tube length. The flux tubes then draw the so-called Steiner tree linking the quarks and antiquarks of the system. There may be more than one Steiner point in such cases: Numerical algorithms exist to compute them; see~\cite{Bicu3} for example. 

When gluons are present, the situation is different. Indeed, a constituent gluon may either generate an adjoint flux tube or two fundamental ones. In the two-gluon case, lattice results favor an adjoint string linking both gluons~\cite{Bicu1,cas} rather than two superimposed fundamental strings. In a three-gluon system however, recent lattice results indicate that a triangle configuration in which each gluon is linked to the two others by a fundamental flux tube is preferred to a Y-junction made of adjoint strings~\cite{Bicu2}. This result can be understood as follows from energetic considerations. Let us first consider that each gluon in a three-gluon system, located at the apices of a triangle $ABC$, generates an adjoint flux tube, and that the flux tubes meet at a point $T$. Assuming the Casimir scaling hypothesis, the energy contained in the Y-junction is then $(9/4)\sigma (AT+BT+CT)$. But the triangular inequality leads us to $(9/4)\sigma (AT+BT+CT)>2\sigma(AT+BT+CT)\geq \sigma (AB+BC+CA)$~\cite{Mathieu:2005wc}: The energy of a triangular configuration is energetically more favorable than the Y-junction, as also suggested by the lattice study~\cite{Bicu2}.

\begin{figure}[ht]
	\centering
		\includegraphics*[width=13cm]{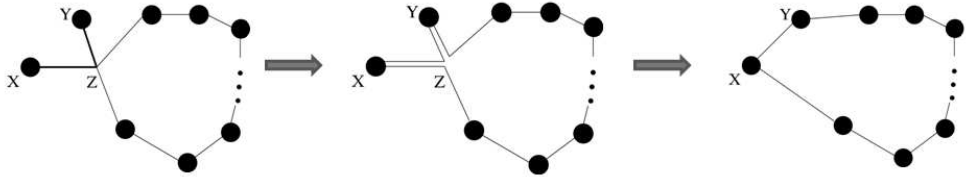}
	\caption{Graphical illustration of the most energetically favorable ${\cal N}$-gluon configuration.}
	\label{Nglu}
\end{figure}

The energetic argument can be extended to a system made of ${\cal N}$ gluons. Let us assume that the most favored configuration of a $({\cal N}-1)$-gluon system is the one in which each gluon is linked to its nearest neighbors by a fundamental flux tube. We know that it is the case for ${\cal N}-1=3$ gluons. Then we choose a pair of gluons, denoted as $X$ and $Y$, that each generate an adjoint flux tube. These flux tubes meet in $Z$, from which a flux tube starts and connect in some way to the ${\cal N}-2$ remaining gluons. $Z$ is allowed to be in the $\bm 8$, $\bm{ 10}$, $\overline{\bm{10}}$, or $\bm{ 27}$ color representations. The less energetic flux tube generated by $Z$ is obtained when $Z$ is in a color octet. Consequently, $Z$ and the ${\cal N}-2$ remaining gluons are equivalent to a $({\cal N}-1)$-gluon system, for which the flux tube configuration is known (closed-chain-like configuration). But at the same time, $XYZ$ is equivalent to a three-gluon system: $X$, $Y$, and $Z$ must be connected by fundamental flux tubes. One finally concludes that the most favored configuration is the one in which $Z$ is absent and in which a given gluon is linked to its nearest neighbors by two fundamental flux tubes. A schematic illustration of the above discussion is shown in Fig.~\ref{Nglu}.

This suggests that a relevant ${\cal N}$-gluon Lagrangian generalizing~(\ref{lag1}) for ${\cal N}>2$ is
\begin{equation}\label{lagn}
		{\cal L}=-m_g\sum^{{\cal N}}_{i=1}\sqrt{\dot{\rm x}^2_i}-\sigma\sum^{{\cal N}}_{i=1} \int^{1}_{0}d\theta\, \sqrt{(\dot{\rm w}_i\, {\rm w}'_i)^2-\dot{\rm w}^2_i\, {\rm w}'^2_i},
\end{equation}
with
\begin{equation}
\rm w_i=\theta \rm x_i+(1-\theta)\rm x_{i+1}\qquad {\rm and}\qquad {\rm x}_{{\cal N}+1}={\rm x}_1.	
\end{equation}
The resulting Hamiltonian equations are quite complex, see for example~\cite{bramb95} in the case of a Y-junction. Let us work again in the instantaneous gauge, and then neglect the contribution of the flux tubes to the momentum of the system. With this approximation one is led to the spinless Salpeter Hamiltonian
\begin{equation}
	H_0^{({\cal N})}=\sum^{{\cal N}}_{i=1}\sqrt{\vec p^{\, 2}_i+m^2_g}+\sigma \sum^{{\cal N}}_{i=1}|\vec x_i-\vec x_{i+1}|.
\end{equation}
It is worth saying that not only this last Hamiltonian has to be considered in order to make explicit computations, but also the nontrivial symmetries of the wave function induced in the various $J^{PC}$ channels by the color wave function (commanding the charge conjugation) and the Pauli principle. These symmetries can become nontrivial as soon as ${\cal N}\geq 3$~\cite{glub}.

\section{Regge trajectories}\label{regge}

\subsection{Perturbative flux tube and spinless Salpeter Hamiltonian}\label{pftssh}
So far we have presented three different types of Hamiltonians that can be used in hadronic physics. The first one is the relativistic flux tube Hamiltonian~(\ref{tf_equa1}), that contains the full relativistic dynamics encoded in Lagrangian~(\ref{lag1}), itself derived from QCD. Numerical computations become simpler if one is able to deal with a closed Hamiltonian; that is why the perturbative flux tube model, defined by (\ref{ssh}) and (\ref{pft}), can be used as an approximation of the full relativistic flux tube model. Finally, the dynamical contribution of the flux tube may completely be neglected and the resulting Hamiltonian is the spinless Salpeter one~(\ref{ssh}).

In this section, we compare the three frameworks in the ultrarelativistic limit, \textit{i.e.} $m_g=0$, that is the case in which the relativistic effects are the most important. It is well-known that in the case of massless particles, all those approaches lead to particular orbital (Regge) and radial trajectories: Defining $\left\langle \vec L^{\, 2}\right\rangle=\ell(\ell+1)$, it is observed that $M^2\propto \ell$ and $M^2\propto n$ for large $\ell$ and $n$ respectively. Notice that, in the meson case, $\ell$ is the orbital angular momentum of the system, while $\ell(\ell+1)=J(J+1)+2\lambda_1\lambda_2$ for the glueballs. Let us now give a more quantitative description of that property.

\begin{figure}[t]
	\centering
		\includegraphics*[width=9cm]{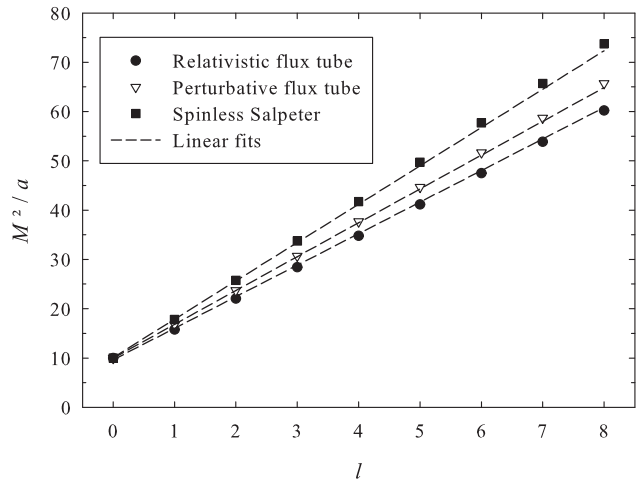}
	\caption{Squared masses in units of the string tension $a$ versus $\ell$ for $n=0$. A comparison is made between the results of the relativistic flux tube model (circles), the perturbative flux tube model (triangles), and the spinless Salpeter Hamiltonian (squares). The linear fits~(\ref{fit1})-(\ref{fit3}) are also plotted. Calculations are made in the ultrarelativistic limit $m_g=0$.}
	\label{fig1}
\end{figure}

The only energy scale in the massless case is $\sqrt a$. Consequently, $M^2/a$ is a dimensionless quantity that can be computed numerically. Our procedure is the following. Thanks to the Lagrange mesh method, which is a simple and accurate numerical technique, we are able to solve the relativistic flux tube model~\cite{buis041} and spinless Salpeter-type Hamiltonians~\cite{sem01}. Then, we compute the square masses for each approach and for $0\leq \ell,n\leq 8$, which is enough to clearly observe the expected linear trajectories. As an example, we plot in Fig.~\ref{fig1} the $n=0$ Regge trajectory computed in the three considered approaches. The linear behavior is clearly observed, but the Regge slope is different in each case. Fits on the complete sets of data that we have computed show that the square masses are in excellent agreement with the following linear forms:
\begin{equation}\label{fit1}
M^2=2\pi a(1.944\, n+1.238\, \ell+1.606)
\end{equation}
for the spinless Salpeter Hamiltonian~(\ref{ssh}),
\begin{equation}\label{fit2}
M^2=2\pi a(1.968\, n+1.092\, \ell+1.586)
\end{equation}
for the perturbative flux tube model~(\ref{ssh})-(\ref{pft}), and
\begin{equation}\label{fit3}
M^2=2\pi a(2.003\, n+1.018\, \ell+1.533)\approx2\pi a (2n+\ell+3/2)
\end{equation}
for the relativistic flux tube model~(\ref{tf_equa1})-(\ref{tf_equa2}). The qualitative behavior of the mass spectrum in all those three approaches is clearly identical. That is why, even if a spinless Salpeter Hamiltonian is a crude approximation of the relativistic flux tube model, it can reproduce the same data provided that the parameters are correctly fitted. Indeed, let us imagine that we want to reproduce the Regge trajectories of light mesons. Since the Regge slopes (in units of $a$) are quite different, 7.779, 6.861, and 6.396 for equations~(\ref{fit1})-(\ref{fit3}), one has to choose a different value for the string tension in each case in order to agree with experiment.

It is also worth mentioning that the radial and Regge slopes are never equal, even approximately. It turns out from the fit~(\ref{fit3}) that the relativistic flux tube model actually exhibits a remarkable harmonic oscillator degeneracy, with a radial slope twice as large as the Regge slope. It implies in particular that, when applied to light mesons, the flux tube model alone is unable to explain the Coulomb-like degeneracy that seems to be experimentally observed in the highly excited mass spectrum~\cite{afon}. Notice that the slope of radial trajectories is almost insensitive to the various approximations, while the Regge slope varies significantly. This is in agreement with~\cite{olsrad}, where it is shown that the relativistic flux tube model is equivalent to a spinless Salpeter Hamiltonian for large radial excitations.

\subsection{Open-closed flux tube duality}\label{dual}
It is readily checked that the mass spectrum~(\ref{fit3}) corresponds not only to the relativistic flux tube model but also to the following Hamiltonian
\begin{equation}
	H_c=\sqrt{\vec p^{\, 2}+(2\pi\tilde a\, r)^2}\, ,\qquad {\rm with}\qquad \tilde a=\frac{a}{2}.
\end{equation}
This last Hamiltonian has a nonstandard form, which actually corresponds to a closed circular flux tube of energy density $\tilde a=a/2$ and of radius $r\gg1/\sqrt{\tilde a}$~\cite{closed}. Consequently, the following models lead to equivalent mass spectra: An open flux tube of energy density $a$ and length $r$ with massless particles at his ends and a closed circular flux tube of energy density $a/2$ and radius $r$.

Such a duality is particularly interesting in the case of glueballs since both the closed- and open-flux tube approaches can be found in the literature. The open flux tube framework has the advantage of providing a common description of light mesons and glueballs~\cite{expe}. But, it is sometimes argued that gluons in a glueball should all have the same status and not being either constituent or part of a flux tube. That is why some authors have proposed that glueballs should be described by closed fundamental flux tubes~\cite{latti2,closed,Isgur}. Such a picture has also the advantage of being closer to the lattice QCD formulation, in which pure glue states are described by loops on the lattice~\cite{Isgur}. In the closed flux tube approach, the different $J^{PC}$ states are generated by orbital or phononic excitations of the flux tube. It is worth noting that, if we set $a=\sigma_a$ for the open flux tube in a glueball, $a/2$ is not far from the expected value $\sigma$.

For completeness we can quote that more exotic, knotted, configurations of the flux tubes have been suggested to be relevant in the modelization of glueballs~\cite{knot}, and that a description of glueballs as two-dimensional spherical membranes has also been proposed in~\cite{gaba}. In this last reference, the effective Hamiltonian reads $\sqrt{\vec p^{\, 2}+(4\pi T r^2)^2}$, the potential term being the surface of the spherical membrane of tension $T$. The squared mass operator is in this last case a quartic oscillator, whose spectrum would be of the form $M^2/T^{2/3}\propto (2.27 n+\ell+1.68)^{4/3}$~\cite{afm}, in disagreement with the open flux tube behavior.

The ${\cal N}$-gluon generalization~(\ref{lagn}) of the flux tube model is nothing else than a closed fundamental flux tube with transverse constituent gluons attached on it. That picture seems roughly compatible with an open- closed-flux tube duality, with the dynamics of the constituent gluon seen as an alternative way of modeling the various phononic or orbital excitations of a pure closed flux tube. However, we point out that the celebrated closed flux tube model of Isgur and Paton~\cite{Isgur} is not in good agreement with recent lattice data when taken in its original form~\cite{ipat}. In particular, it has difficulties to reproduce the important splittings between $C=+$ and $-$ glueballs with the same $J^P$. It thus appears that improvement of this model are needed in order to understand the lattice QCD glueball spectrum.

\section{The glueball spectrum}\label{gluspec}

The relativistic flux tube model presented in section~\ref{ftm} provides us with a relativistic -- although not explicitly covariant -- description of hadronic bound states. Supplemented by other physical mechanisms that we present hereafter, it can be used to understand the lattice data.

\subsection{Short range interactions}
The flux tube stands for the long-range interactions, \textit{i.e}. the confinement. At short distances however, other physical mechanisms become dominant. These are mainly given by the one-gluon-exchange processes between color sources. The corresponding short-range potentials can be computed from the QCD Feynman diagrams at tree level or from the perturbative part of the Wilson loop. In the quark-antiquark case, one is led to the well-known Fermi-Breit interaction. Other cases such as gluon-gluon and quark-gluon potentials have been investigated for example in~\cite{oge}. As expected, it appears that, at the dominant order, the short-range potential is always given by a Coulomb term scaled by some color factor depending on the considered color sources; in particular one is led to
\begin{eqnarray}\label{ggpot}
    \Delta H_{oge}=-3\frac{\alpha_S}{r}
\end{eqnarray}
for a two-gluon glueball, $\alpha_S$ being the strong coupling constant and $r$ the distance between the two gluons. In a three-gluon system, the 3 factor is replaced by $3/2$ because a given gluon pair is in a color octet. In the special case of the four-gluon $0^{+-}$ state, gluon pairs are found to be either in a color decuplet or antidecuplet, leading to the vanishing of $\Delta H_{oge}$~\cite{glub}.

Already at one-loop, $\alpha_S$ is running with the energy scale $\rm q^2$; we know that $\alpha_S(\rm q^2\rightarrow\infty)=0$. Moreover, we consider that $\alpha_S(\rm q^2\rightarrow0)=\alpha_0$. This is coherent with lattice QCD calculations, showing that the static potential between a quark and an antiquark is accurately fitted by the form $\sigma r-(4/3)\alpha_S/r$~\cite{balirep}. An intuitive justification of this last relation is that the total energy of the system is separated into a dominant, nonperturbative, part (the flux tube), and a residual, perturbative, part (the one-gluon-exchange potential). This is in agreement with the aforementioned lattice works~\cite{ichie,taka1,taka2,Bissey,Bicu1,Bicu2,oki1,oki2}, in which the potential energy between the various sources considered is always compatible with the expected one-gluon-exchange term at short distances, while the flux-tube contribution comes into play at long distances. 

\subsection{Instanton-induced forces}

The influence of instantons on light quarks and light mesons has been intensively studied (see~\cite{instanton3} for a complete review about instantons in QCD). It is known, in particular, that instanton-induced forces in the lightest mesons can be included in potential quark models as an isospin-dependent term whose contribution is nonzero in the pseudoscalar channel only~\cite{instanton}. Such a term is needed to reproduce the very low value of the pion mass without lowering the masses of the other mesons.

Unlike the meson case, the effects of instantons on glueballs have been less intensively studied. In a pioneering work~\cite{scha}, it has been computed that direct instantons induce a strong attractive force in the scalar glueball channel and a strong repulsive force in the pseudoscalar one, while no effect is observed in the tensor channel and presumably in the other channels. It is moreover tempting to deduce from~\cite{inst2} that these forces in the scalar and pseudoscalar channels are of equal magnitude but of opposite sign. This has led us to propose the following phenomenological instanton mass term in Ref.~\cite{gluh1}:
\begin{equation}\label{hidef}
	\Delta H_I=-P\, {\cal I}\, \delta_{J,0},
\end{equation}
where $P$ is the parity operator and where the arbitrary parameter ${\cal I}$ is assumed to be positive and constant in first approximation. It is relevant to include instanton-induced forces in glueballs because they play an important role in the meson case and, since they are needed in glueballs as well~\cite{scha,inst2}, it is coherent to include them in the model.

\subsection{Two-gluon spectrum}

Using the Lagrange mesh method~\cite{buis041}, we are now able to compute the two-gluon glueball mass spectrum from (\ref{tf_equa1}) and (\ref{tf_equa2}) in which $H$ is replaced by $H+\Delta H_{oge}+\Delta H_I$ and in which $\langle \vec L^{\, 2}\rangle$ is computed thanks to (\ref{lsdef}), coming from the helicity formalism. The corresponding $J^{PC}$ states have all a positive charge conjugation and are given by~(\ref{glustate}). 

The parameters remain to be fixed. We set $m_g=0$ in the spirit of~\cite{gluh1}, which was a first attempt to include the transversality of the constituent gluons in a potential model. We think that it is interesting for the present model parameters not to be in disagreement with~\cite{expe}, in which the experimental mass spectra of light an heavy mesons are quite satisfactorily reproduced within the perturbative flux tube model. We thus keep the value $\alpha_S=0.4$, used in this last reference in the light meson sector. The value used for the string tension was $0.185$~GeV$^2$ in Ref.~\cite{expe}, ensuring a correct reproduction of the experimental mesonic Regge slopes. Let us denote $\sigma_{pft}=0.185$~GeV$^2$ the fundamental string tension fitted on light mesons in the perturbative flux tube model~\cite{expe}. Then the same Regge slope will be obtained by using  $\sigma=(6.863/6.398)\sigma_{pft}=0.198~{\rm GeV}^2$ in the relativistic flux tube model. In summary we can set, in the two-gluon relativistic flux tube model
\begin{equation}\label{params}
m_g=0,\	\sigma_a=(9/4)\sigma,\ \sigma=0.2~{\rm GeV}^2,\ \alpha_S=0.4,\ {\rm and}\ {\cal I}=0.450~{\rm GeV},
\end{equation}
where the magnitude of instanton-induced forces is fitted on the lattice splitting between the scalar and pseudoscalar states.

The mass spectrum we obtain by computing two-gluon glueball masses within the relativistic flux tube model is given in Table~\ref{tab1}. It can be seen that the agreement with the states that are currently known in lattice QCD is good, all the computed masses falling into the error bars. Only the $3^{++}$ state is missed; we indeed find a very low value for it, so low that the lattice mass rather corresponds to the first radially excited state. To our knowledge, there is no theoretical reason to exclude the $3^{++}$ state with $n=0$. Since that glueball is the only one coming from the $D_-$ family, it would be interesting to compute the $5^{++}$ glueball mass on the lattice and see whether it agrees more with the $n=0$ or $n=1$ state in our model.

\begin{table}
\centering	\setlength{\extrarowheight}{2pt}
		\begin{tabular}{clcc}
		\hline\hline
			$J^{PC}$ & Lattice (GeV) & Model (GeV) & State \\
			\hline
 $0^{++}$  & 1.710$\pm$130~\cite{latti1}& 1.710 & $\left|S_+;0^+\right\rangle$\\
           & 2.670$\pm$310~\cite{latti0}& 2.631 &$\left|S_+;0^+\right\rangle$\\
 $0^{-+}$  & 2.560$\pm$155~\cite{latti1}& 2.610& $\left|S_-;0^-\right\rangle$\\
           & 3.640$\pm$240~\cite{latti0}& 3.531& $\left|S_-;0^-\right\rangle$ \\
 $2^{++}$  & 2.390$\pm$150~\cite{latti1}& 2.529& $\left|D_+;2^+\right\rangle$\\
           &                         & 2.972& $\left|S_+;2^+\right\rangle$\\
 $2^{-+}$  & 3.040$\pm$190~\cite{latti1}& 2.972 & $\left|S_-;2^-\right\rangle$\\
           & 3.890$\pm$230~\cite{latti1} & 3.756 &  $\left|S_-;2^-\right\rangle$\\
 $3^{++}$  & 3.670$\pm$230~\cite{latti1} & 3.132 &$\left|D_-;3^+\right\rangle$\\
           &                                   & 3.893&$\left|D_-;3^+\right\rangle$ \\
 $4^{++}$  & 3.650$\pm$240~\cite{latti3} & 3.599 &$\left|D_+;4^+\right\rangle$\\
           &                          & 3.775   &$\left|S_+;4^+\right\rangle$\\
 $4^{-+}$  &                                  & 3.775 &$\left|S_-;4^-\right\rangle$ \\
 $5^{++}$  &                                  & 3.999  &$\left|D_-;5^+\right\rangle$\\
           &                                  & 4.655  &$\left|D_-;5^+\right\rangle$\\
		\hline	\hline
		\end{tabular}
	\caption{Comparison between the $C=+$ glueball masses computed in pure gauge lattice QCD~\cite{latti0,latti1,latti3} and the spectrum of the relativistic flux tube model with parameters~(\ref{params}). The corresponding helicity states are given in the last column.}
	\label{tab1}
\end{table}

It is worth pointing out that the masses of the states $\left|S_+;(J>0)^+\right\rangle$ and $\left|S_-;(J>0)^-\right\rangle$ are equal in our approach because of~(\ref{lapp}). It was already suggested in the pioneering work~\cite{barnes} that such a characteristic pattern would be an interesting tool to identify glueball resonances. Moreover, this degeneracy cannot be lifted by the introduction of tree-level spin-dependent operator like spin-orbit, spin-spin, \textit{etc.} operators~\cite{barnes,gluh1}. A mixing term between the $S_+$- and $D_+$-states would be able to affect this mass degeneracy. The existence of such a mixing can be understood as follows. Let us consider two transverse gluons on mass shell, with their helicity denoted by $+$ or $-$. If ${\cal A}(\lambda_1,\lambda_2;\lambda_1',\lambda_2')$ is the gluon-gluon scattering amplitude with $\lambda_i$ and $\lambda_i'$ the incoming and outcoming helicities respectively, then it can be shown that, at tree level~\cite{mix}, ${\cal A}(++;++)\neq0$ and ${\cal A}(++;--)\neq0$, but ${\cal A}(++;+-)={\cal A}(++;-+)	=0$. In other words, the helicity-conserving amplitudes are nonzero at tree-level and give rise at the dominant order to the Coulomb term~(\ref{ggpot}). However, the scattering amplitudes of two gluons is zero when only one helicity is reversed. Reversing a single helicity transforms a helicity-singlet in a helicity-doublet and vice-versa. It is thus logical that we obtain no singlet-doublet mixing with tree-level relativistic corrections. At the next order however, scattering amplitudes such as ${\cal A}(++;+-)$ become nonzero and a singlet-doublet mixing can appear. However, the explicit form of the total scattering amplitude beyond tree level is very complex (see Ref.~\cite{mix}), and computing an effective potential from all the allowed diagrams is a task which is beyond the scope of this paper. Another consequence of such a mixing is that it decreases the mass of the lowest-lying $(2k+2)^{++}$ states, improving in particular the agreement of the $2^{++}$ state with the optimal lattice value.

The glueball spectrum of Table~\ref{tab1} has been computed by considering massless constituent gluons. It is worth mentioning that other frameworks like Coulomb gauge QCD rather use a nonzero gluon mass of typically 600~MeV~\cite{couglu}. We have checked that a mass spectrum in agreement with lattice QCD can be obtained with $m_g=0.6$~GeV, $\sigma=0.135$~GeV$^2$, $\alpha_s=0.45$, and ${\cal I}=0.45$~GeV, but the global agreement is less satisfactory than with parameters~(\ref{params}). Those values are inspired from a Coulomb gauge study of hybrid mesons~\cite{general}. 

Finally, it is worth mentioning that the auxiliary field method allows to find an approximate analytical mass formula for the Hamiltonian $2\sqrt{\vec p^{\, 2}+m^2_g}+\sigma_a r-3\alpha_S/r-P\, {\cal I}\, \delta_{J,0}$, which is a simplification of the full relativistic flux tube model~\cite{afmsr}. Correcting the formula obtained by taking into account the result~(\ref{fit3}), the final mass formula, for $m_g^2\ll \sigma_a$, reads~\cite{afmsr}
\begin{equation}\label{2gma}
	M_{2g}=M_0+\frac{4M_0m^2_g}{M_0^2+3\pi\sigma_a\alpha_S}-P\, {\cal I}\, \delta_{J,0},
\end{equation}
where
\begin{equation}\label{2gmabis}
	M_0=\sqrt{\pi\sigma_a(4n+2\ell+3(1-\alpha_S))}
\end{equation}
and where $n$ and $\ell$ are the radial and orbital quantum numbers of the gluon pair. $\ell$ is computed through the relation $\ell(\ell+1)=J(J+1)+2\lambda_1\lambda_2$ [see Eq.~(\ref{lsdef})]. This last formula reproduces the glueball spectrum for two massless gluons computed within the relativistic flux tube model with a global accuracy better than 4\% (but 9\% at $J=0$).

\subsection{Three- and four-gluon systems}\label{34g}

It is possible to obtain an estimation of the lightest three-gluon glueball mass thanks to analytical results obtained with the auxiliary field method in~\cite{afm2}. It is shown in this last reference that an accurate approximate analytical mass formula for the Hamiltonian $H_{3g}=\sum^3_{i=1}\sqrt{\vec p^{\, 2}_i}+\sum^3_{i<j=1}\left(\sigma |\vec x_i-\vec x_j|-(3/2)\alpha_S/|\vec x_i-\vec x_j|\right)$, intended to describe glueballs made of three gluons, is given by
\begin{equation}\label{mggg}
	M^2_{3g}=(12\sqrt 3\, N-54\alpha_S)\sigma,\qquad {\rm with}\qquad N=3+\sum^2_{i=1}(2n_i+\ell_i),
\end{equation}
where $n_i$ and $\ell_i$ denote the radial and orbital quantum numbers associated to the two Jacobi coordinates. This last formula is actually an (accurate) upper bound of the exact mass spectrum~\cite{afm2}. We also point out that it has been obtained within a spin-independent formalism. In other words, only the orbital angular momenta are present, and no input coming from the gluons helicity has been added. The ground state of~(\ref{mggg}) has a mass equal to $(36\sqrt 3-54\alpha_S)\sigma$, that we expect to be an underestimation of the ground-state mass. Indeed, that value is obtained by setting all the relative momenta equal to zero, \textit{i.e.} $N=0$, and it can be seen, already in two-gluon case, that the introduction of transverse degrees of freedom adds components of higher orbital momenta in the system, tending to increase the total mass~\cite{gluh1}. With the parameters~(\ref{params}) we obtain $M_{3g}=2.855$~GeV; since the lightest $C=-$ glueball computed on the lattice has a mass of $2.980\pm170$~GeV, it appears to be a relevant lower bound.

We proposed in Ref.~\cite{glub} to see the $0^{+-}$ glueball as a low-lying four-gluon state rather than a highly-excited three-gluon state. Let us check that hypothesis. The particular color structure of this state (see section~\ref{fourglu}) causes the short-range interactions to vanish, and enforces a particular mixed symmetry that suggests an approximate way of computing the $0^{+-}$ mass by using two-gluon clusters (see Ref.~\cite{glub} for more details). First, we compute the mass of the lightest antisymmetric state made of two gluons linked by a fundamental flux tube of tension $\sigma$. This state is a $J=1$ one, for which $\left\langle \vec L^{\, 2}\right\rangle=4$ following~(\ref{lsdef}), and the relativistic flux tube model gives a cluster mass of $M_c=1.964$~GeV. Second, we couple both massive clusters in a $S$-wave channel, which is the lightest symmetric configuration of these clusters, linked by a flux tube of energy density $2\sigma$. The final mass is $4.918$~GeV, a value compatible with the lattice one of $4.780\pm0.290$~GeV~\cite{latti1}.

\section{Glueball decay widths}\label{decay}
Together with the mass, the total decay width is a particularly relevant observable characterising a given hadron. We adapt in this section the method of~\cite{abreu}, giving a way to estimate the two-gluon glueball decay widths within a flux tube approach.

\subsection{General mechanism}

Let us consider a two-body hadron (meson or glueball typically). The flux tube linking both constituents is a particular configuration of the gluonic field in QCD; it is purely chromoelectric in the rest frame. Within this chromoelectric field, a particle-antiparticle pair can be created from the vacuum thanks to the Schwinger effect~\cite{sche}. This creation eventually leads to a breaking of the flux tube, and thus to the decay of the considered system. In the case of a meson, say $Q\bar Q$, a quark-antiquark pair $q\bar q$ can be created inside the flux tube and causes the meson to decay as $Q\bar Q\rightarrow q\bar Q+\bar q Q$. It is known for a long time that such a mechanism can successfully describe the decay widths of mesons, and even baryons~\cite{decay}. The idea can be generalized to the glueball case, with the creation of a quark-antiquark pair within the adjoint flux tube. Other processes can also occur in glueballs: Simultaneous creation of two quark-antiquark pairs within the flux tube, quark-antiquark disintegration of a constituent gluon, \textit{etc}. Many processes are actually allowed because more energy is stored in an adjoint flux tube than in a fundamental one, and glueball decays are clearly more complex than meson ones. That is why the calculations we will perform here, which are based on the $q\bar q$ creation picture, only aim to give rough estimations of the glueball total decay widths. We remark that this approach should not be used in the case of heavy quarkonia since decays induced by short-range processes may become dominant with respect to the flux-tube induced ones.

The quantity $\sqrt{\left\langle r^2\right\rangle}$ is a measure of the flux-tube length. It is plausible that the probability to produce a quark-antiquark pair is proportional to the string length since this production is linked to the energy carried by the chromoelectric field. The energy scale of the flux tube model with string tension $a$ is given by $\sqrt a$ in the ultrarelativistic limit. We are thus led, in agreement with~\cite{abreu}, to the following ansatz for the decay width
\begin{equation}\label{decayqq}
	\Gamma=\gamma\, \sqrt{a}(\sqrt{\left\langle x^2\right\rangle}-x_0),
\end{equation}
where $\vec x=\vec r \sqrt{a}$ is used together with $\vec q=\vec p/\sqrt{a}$ to get dimensionless average values, and where $\gamma$ and $x_0$ are assumed to be universal constants. Notice that (\ref{decayqq}) could be written on the alternative form $\Gamma=\bar\gamma(M-\bar x_0)$, where the hadron mass explicitly appears.

\begin{figure}[ht]
	\centering
		\includegraphics*[width=9cm]{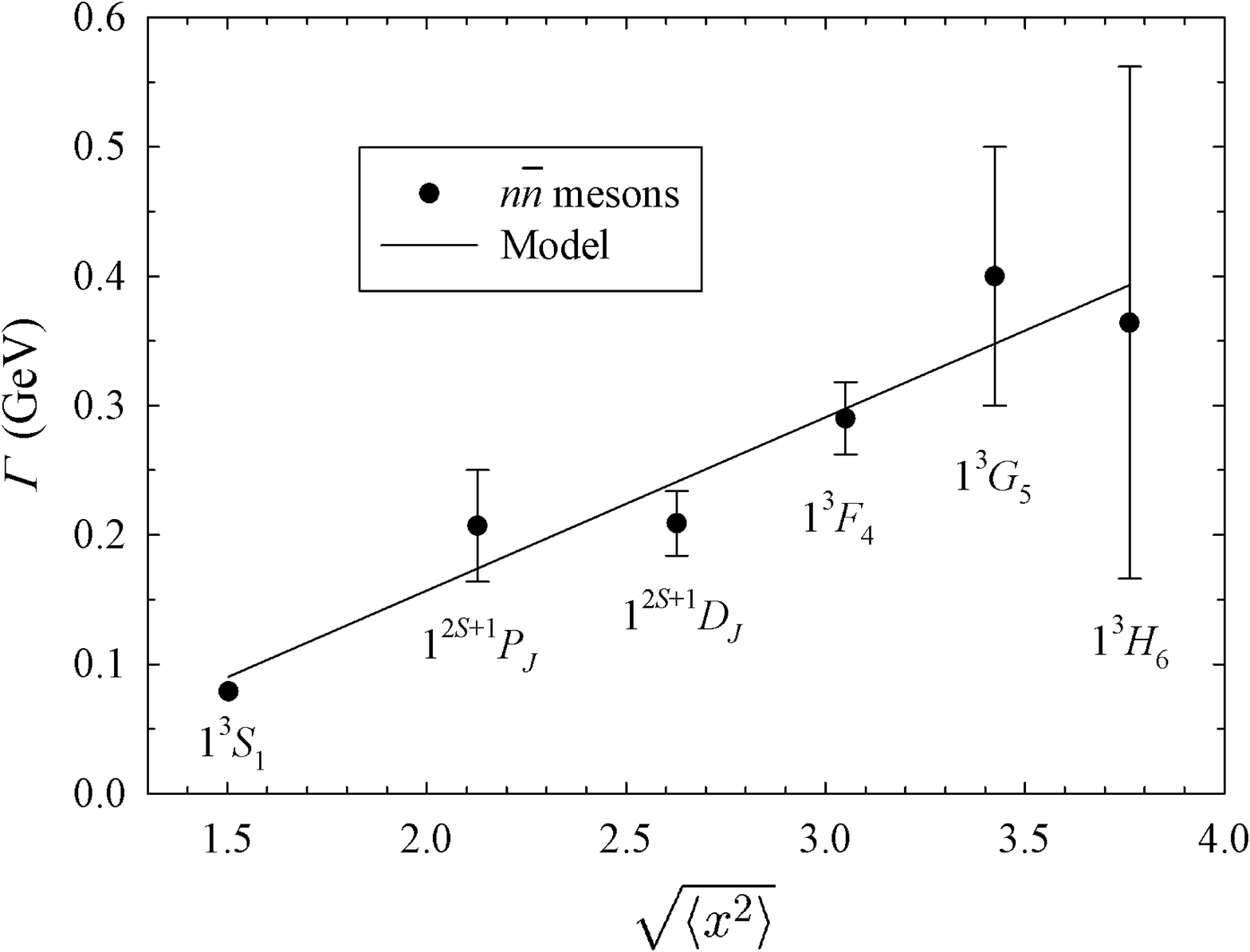}
	\caption{Plot of the spin- and isospin-averaged experimental decay widths of $n\bar n$ mesons sharing the same spectroscopic assignment $(n+1)^{2S+1}L_J$ versus $\sqrt{\langle x^2\rangle}$ (circles). The form~(\ref{decayqq}) with the parameters~(\ref{fitde}) and $a=\sigma$ is plotted for comparison (solid line).}
	\label{figd}
\end{figure}

\subsection{Numerical results}

One has first to fit the parameters $\gamma$ and $x_0$. To do that, we resort to the available experimental data concerning light mesons~\cite{PDG}, that we denote as $n\bar n$ mesons, $n$ standing for $u$ and $d$ quarks. The present approach is spin- and isospin-independent. Consequently, we have to apply it to the spin- and isospin-averaged decay widths of the lightest mesons, as done in~\cite{abreu}. Only the lightest states are taken into account because they are expected to possess the largest radii, thus to be the ones for which flux-tube induced effects are the most relevant. 
The values $\sqrt{\langle x^2\rangle}$ are computed by using the relativistic flux tube model for the quark-antiquark pair with the Coulomb term $-(4/3)\alpha_S/x$ and the parameters~(\ref{params}). The couples $(\sqrt{\langle x^2\rangle},\Gamma)$ are then fitted to~(\ref{decayqq}) with $a=\sigma$, leading to
\begin{equation}\label{fitde}
\gamma=0.300\pm0.045,\qquad x_0=0.827\pm0.309.
\end{equation}
The result in the light meson sector is plotted in Fig.~\ref{figd}.

\begin{table}[ht]
	\centering	\setlength{\extrarowheight}{2pt}
		\begin{tabular}{ccc}
	\hline\hline
$J^{PC}$ & $\Gamma_{gg}$ (MeV) & $M$ (GeV)\\
\hline
$0^{++}$ &  219$\pm$95 & 1.710\\
$0^{-+}$ &  219$\pm$95 & 2.610\\
$2^{++}$ &  286$\pm$105 & 2.529\\
$2^{-+}$ &  367$\pm$117 & 2.972\\
$3^{++}$ &  397$\pm$122 & 3.132\\
$4^{++}$ &  484$\pm$135 & 3.599\\
$4^{-+}$ &  517$\pm$140 & 3.775\\
			\hline\hline
	\end{tabular}
	\caption{Total decay width of some low-lying glueballs computed with formula~(\ref{decayqq}), the parameters~(\ref{fitde}), and $a=\sigma_a$. The error bars on $\Gamma_{gg}$ come from the uncertainty on $\gamma$ and $x_0$. The corresponding masses are recalled in the last column (see Table~\ref{tab1}). }
	\label{tab5}
\end{table}

We can then estimate the total decay widths of some glueballs by computing the corresponding values of $\sqrt{\langle x^2\rangle}$. Some results are shown in Table~\ref{tab5}. The decay widths of the scalar and pseudoscalar glueballs are degenerate in our approach, since instanton-induced interactions are only added as a constant term and do not affect the radius. It is interesting to compare these results with some experimental data. It has been suggested in several works that the $f_0(1710)$ could be a mainly glueball state~\cite{mix1}. Its experimental mass and decay width are respectively $1.724\pm0.007$~GeV and $137\pm8$~MeV~\cite{PDG}. The decay width is lower than our estimate, although still falling in the error bars of our prediction. It is worth mentioning the detailed Coulomb gauge computation of~\cite{cgglu} leading to the quite low value of $100$ MeV for the scalar glueball decay width. Those results could favor the glueball interpretation of the $f_0(1710)$, although this last work was rather concerned with the $f_0(1810)$ resonance. There are two tensor states which are located in the mass range of lattice QCD: The $f_2(2300)$ with a mass and decay width of $2.297\pm0.028$~GeV and $149\pm40$~MeV, and the $f_2(2340)$ with a mass and decay width of $2.339\pm0.060$~GeV and $319^{+80}_{-70}$~MeV respectively~\cite{PDG}. The masses of these states are compatible, but their decay widths are not. Our results rather favor the $f_2(2340)$ as a tensor glueball candidate.

\section{The pseudoscalar glueball}\label{etaetap}
Recently, the $\eta$-$\eta'$ mixing angle as well as the gluonium content of the $\eta'$ have been computed by the KLOE collaboration from a study of the channels $\phi(1020)\rightarrow\eta\gamma$ and $\eta'\gamma$~\cite{kloe}. Let us recall the results obtained in these last references. First of all, the $\eta$ and $\eta'$ wave functions are are decomposed into three Fock-space components: $\left|n\bar n\right\rangle=(\left|u\bar u\right\rangle+\left|d\bar d\right\rangle)/\sqrt 2$, $\left|s\bar s\right\rangle$ and a glueball component that we denote $\left|gg\right\rangle $. With the following parameterization,
\begin{eqnarray}\label{paramstate}
\left|\eta\right\rangle&=&\cos\varphi_p\left|n\bar n\right\rangle-\sin\varphi_p\left|s\bar s\right\rangle,\nonumber\\
\left|\eta'\right\rangle&=&\cos\varphi_g\sin\varphi_p\left|n\bar n\right\rangle+\cos\varphi_g\cos\varphi_p\left|s\bar s\right\rangle+\nonumber\\
&&
+\sin\varphi_g\left|gg\right\rangle,
\end{eqnarray}
one obtains from a fit to the experimental data~\cite{kloe}
\begin{eqnarray}\label{resk}
\varphi_p=(40.0\pm0.7)^{\rm o},\qquad Z^2_g&=&\sin^2\varphi_g=0.13\pm0.04\nonumber\\
&&\Rightarrow \varphi_g=(21.1\pm3.4)^{\rm o}.
\end{eqnarray}
It is assumed that the $\eta$-resonance has no glueball component, while the gluonic content of the $\eta'$ is found to be nonzero.

The parameterization~(\ref{paramstate}) allows the existence of a third state, orthogonal to the $\eta$ and $\eta'$ quantum states, expressed as
\begin{equation}\label{psglu}
	\left|G\right\rangle=-\sin\varphi_g\sin\varphi_p\left|n\bar n\right\rangle-\sin\varphi_g\cos\varphi_p\left|s\bar s\right\rangle+\cos\varphi_g\left|gg\right\rangle.
\end{equation}
Introducing rotation matrices, it is readily checked that
\begin{equation}\label{eigen}
\begin{pmatrix}
\left|\eta\right\rangle\\ \left|\eta'\right\rangle \\ \left|G\right\rangle
\end{pmatrix}	
=
\begin{pmatrix}
1 & 0 & 0\\
0 & \cos \varphi_g & \sin\varphi_g\\
0 & -\sin\varphi_g & \cos\varphi_g
\end{pmatrix}
\begin{pmatrix}
\cos \varphi_p & -\sin\varphi_p & 0\\
\sin\varphi_p & \cos\varphi_p & 0\\
0 & 0 & 1
\end{pmatrix}
	\begin{pmatrix}
\left|n\bar n\right\rangle\\ \left|s\bar s\right\rangle \\ \left|gg\right\rangle
\end{pmatrix}	\equiv \textbf{U}\begin{pmatrix}
\left|n\bar n\right\rangle\\ \left|s\bar s\right\rangle \\ \left|gg\right\rangle
\end{pmatrix}.
\end{equation}
Let us now define $M={\rm diag}(M_\eta,M_{\eta'},M_G)$. Then, the states~(\ref{eigen}) are eigenstates of the mass matrix
\begin{equation}
	\tilde M=\textbf{U}^\dagger\, M\, \textbf{U}.
\end{equation}
The diagonal elements of $\tilde M$ are the masses of the unmixed pseudoscalar $\left|n\bar n\right\rangle$, $\left|s\bar s\right\rangle$, and $\left|gg\right\rangle$ states; we denote $M_{gg}$ this last element. $M_{gg}$ can be {\it a priori} different from the pure gauge pseudoscalar glueball mass since quark loops inside the correlator may change its value. A possible effect of sea quarks is to decrease glueball masses (see for example Ref.~\cite{review} for more information). In \cite{Hart:2001fp} the scalar glueball mass with $N_f=2$ is found to be 15$\%$ lower than the pure gauge mass. But no definitive conclusions can be drawn at this stage (see Ref.~\cite{Bali:2000vr}, in which non significant difference with the pure gauge mass is found).

>From the results~(\ref{resk}), $\left|G\right\rangle$ appears as being a glueball at 87\%; it could actually correspond to the ``physical" pseudoscalar glueball. Its mass can be estimated from the KLOE data. Some algebra indeed shows that the following relation holds
\begin{equation}\label{pseudom}
	M_G=M_{gg}+(M_{gg}-M_{\eta'})\, \tan^2\varphi_g.
\end{equation}
Formula~(\ref{pseudom}) gives the mass of the physical pseudoscalar glueball in terms of the experimentally measured mixing angle $\varphi_g$ and the bare pseudoscalar glueball mass. Taking into account the error on the determination of $\varphi_g$~\cite{kloe}, we find
\begin{equation}
	M_G=(1.149\pm0.056)M_{gg}-(0.143\pm0.053)\ {\rm GeV},
\end{equation}
where $M_{\eta'}=957.66\pm0.24$~MeV~\cite{PDG}.
Using our pure gauge value $M_{gg}=2.610$~GeV, we are eventually led to suggest the existence of a pseudoscalar glueball candidate with a mass of $M_G=2.856\pm0.199$~GeV and a total decay width around 220-315~MeV (we take the upper values of Table~\ref{tab5} because of the repulsive instanton interactions and because the mesonic components allow for more decay processes than in the case of a bare glueball). Interestingly, such a mass is not far from the $\eta_c$-meson, whose mass is $2980.3\pm1.2$~MeV but whose decay width is only $26.7\pm3.0$~MeV. For completeness, we mention that following a previous work~\cite{koch}, the $\eta_c$-glueball mixing could be large, while there are preliminary lattice results suggesting that the $\eta'$-$\eta_c$ mixing is small~\cite{balieta}. So far it is not known how large is the influence of sea quarks on the pseudoscalar glueball. In particular the $\eta(1405)$ is sometimes considered as a pseudoscalar state with a large glueball component, hence it could correspond to $|G\rangle$. This conclusion is shared by the recent works~\cite{cheng,li}, based on the FKS formalism~\cite{fks}, with the same experimental input as in the present study. But following~(\ref{pseudom}), this would imply $M_{gg}\approx 1.35$~GeV, a value that would demand huge sea quarks effects to be reached.

It is finally worth saying that there is a controversy about the KLOE results~\cite{kloe}. For example, one can quote Ref.~\cite{escri}, in which it is found through another analysis of radiative scalar and pseudoscalar meson decays that $\varphi_p=(41.4\pm1.3)^{\rm o}$ and $\varphi_g=(12\pm13)^{\rm o}$. Both the $\eta$ and $\eta'$ resonances are compatible with a zero gluonium component in this case. Notice however that more recent computations of the same author agree with a nonzero gluonic component in the $\eta'$~\cite{escri2}.

\section{Glueballs and the Pomeron}\label{glupom}
It is well known that in the many high energy reactions with small momentum transfer, the exchange by the
highest-lying Regge trajectory, called as a soft Pomeron, gives the dominant contribution. This exchange carries
vacuum quantum numbers $PC=++$ and has very peculiar properties in comparison with the usual Regge pole trajectories. From the analysis of $pp$ and $p\bar p$ cross sections it follows that the Pomeron trajectory has a linear behavior reading~\cite{pom0}
\begin{equation}\label{pomer}
	\alpha(t)=1.08+0.25\, t.
\end{equation}
It is expected that such a trajectory should contain physical hadron states, with $t=M^2$ and $\alpha_P(t)=J$. But the Pomeron does not seem to be related to light mesons since the latter have lower intercepts and larger slopes. There has been a long-standing speculation that the physical particles on the Pomeron trajectory might be glueballs. In the lattice study~\cite{latti2}, it is shown that the lightest $(2,4,6)^{++}$ pure gauge states lie on a Regge trajectory that is consistent with the Pomeron~(\ref{pomer}). However, the masses that are found in~\cite{latti2} are significantly lower than the lattice studies~\cite{latti0,latti1} on which we rely in the present approach. Furthermore, in a recent calculation, Meyer updated his value for the scalar and the tensor
glueball masses~\cite{meyer08}. Although being obtained with a different method, the two new masses are, in this reference,
in agreement with Refs.~\cite{latti0,latti1}. We thus conclude that this last mass spectrum is a reliable one and that it is worth discussing a bit the compatibility of our results with the Pomeron.

First of all, the linearity of the trajectory~(\ref{pomer}) is in qualitative agreement with the flux-tube picture for glueballs. The lightest glueball states with $J^{PC}=$(even$\geq2$)$^{++}$ are generated by the $D_+$ family. Within our model, it is readily computed that the squared masses of this family, with $J$ running from 2 to 20, can be fitted by the following form
\begin{equation}\label{doublet}
	J=0.34\, M^2-0.41\approx\frac{1}{2\pi\sigma_a}(M^2-1.16).
\end{equation}
 The slope of this trajectory is quite different from 0.25~GeV$^{-2}$; actually we should have taken the large value $\sigma=0.29$~GeV$^2$ to reproduce it. This is due to the relativistic effects contained in the flux tube. Within a spinless Salpeter formalism, for the same string tension, we would have found a slope of $[7.77(9/4)\, 0.2]^{-1}$~GeV$^{-2}=0.29$~GeV$^{-2}$, in better agreement with~(\ref{pomer}). One has thus to be careful when claiming that a potential glueball model reproduces the Pomeron slope: This is strongly model-dependent, and nonstandard values of the string tension can be needed to match the slopes. We also see in~(\ref{doublet}) that the intercept is not close to 1: Our glueball trajectory thus do not correspond to the leading Pomeron trajectory. It is worth mentioning that a large-$J$ expansion of formula~(\ref{2gma}) with $m_g=0$ gives
\begin{equation}
J=\frac{M^2_{2g}}{2\pi\sigma_a}+\frac{3}{2}(\alpha_S-1)+ O\left(M^{-2}_{2g}\right)\approx 0.35\, M^2_{2g}-0.9.	
\end{equation}
The slope is nearly equal to the fitted form, while the intercept is not. With that formula, $J+2$ could agree with the Pomeron trajectory; this is a supplementary reason to consider that our glueball spectrum rather lies on a daugther trajectory.

Let us now quote the work~\cite{pom1}, where not only $pp$ and $p\bar p$ cross sections are fitted within the Pomeron framework, but also $\pi^{\pm}p$ and $K^{\pm}p$ cross sections. The following results are found: With a single (soft) Pomeron, fits are consistent with
\begin{equation}\label{soft}
	\alpha(t)=1.09+0.33\, t.
\end{equation}
  With two Pomerons (soft and hard), fits are consistent with
\begin{equation}\label{shard}
	\alpha_s(t)=1.07+0.30\, t\qquad{\rm and}\qquad \alpha_h(t)=1.45+0.1\, t,
\end{equation}
although the need for a hard Pomeron is not crucial for the quality of the fits~\cite{pom1}. These results lead to a soft Pomeron slope that is in better agreement with our slope, especially~(\ref{soft}). We can also mention that our $2^{++}$ mass is compatible with the point $\alpha_h(2)$, while in Ref.~\cite{pom2} the tensor glueball is found to lie on the soft Pomeron trajectory.

The fit~(\ref{pomer}) would require the tensor glueball to have a mass around 1.92~GeV, while the fit~(\ref{soft}) favors a tensor glueball mass around 1.65~GeV. Both are values far from our results. In conclusion, the conjecture that the lightest (even-$J$)$^{++}$ glueballs lie on the Pomeron trajectory is not favored by our results; we would rather conclude that these glueballs are located on a daughter trajectory. But, we focused here in reproducing pure gauge lattice data, by computing the spectrum of genuine two-gluon states. It is plausible either that the leading Pomeron trajectory contains no pure glueball but rather mixed states (glueball, mesons, tetraquark,\dots), or that the sea quark effects lead to a strong decrease of even the pure glue states. Both mechanisms escape our current understanding. 

\section{Finite-temperature effects}\label{finiT}
The glueball lattice spectrum, given in Table~\ref{tab1} and satisfactorily reproduced by the relativistic flux tube model, was formally obtained at $T=0$. It is well-known however that nontrivial phenomena occur at nonzero temperature. In particular, there exists a critical temperature in QCD, denoted as $T_c$, above which deconfinement takes place, eventually leading to the celebrated quark-gluon plasma. Since the relativistic flux tube model is an effective approach describing the confinement in QCD, it should not be used at temperatures higher than $T_c$. However, it can be expected that, starting from $T=0$ and increasing the temperature, the flux tube will ``weaken" due to thermal effects and then will disappear at $T_c$. One can think of a progressive disorganization of the chromoelectric field because of random thermal fluctuations, as illustrated in Ref.~\cite{wong}. That picture is supported by lattice computations of the static quark-antiquark energy, showing that the dominant effect below $T_c$ is a diminution of the effective string tension, that can be parameterized by~\cite{kar}
\begin{equation}\label{aeff}
	a(T)=a\, \left(1-\frac{T^2}{T^2_c}\right).
\end{equation}
The variation of the Coulomb term can be ignored in a first approach~\cite{wong}. For completeness, we mention that a potential model describing glueballs above $T_c$ can be found in Ref.~\cite{braut}.

The behavior of the glueball spectrum at finite temperature has been first investigated on the lattice in Refs.~\cite{ishi1,ishi2}. The basic conclusion of those works is that the scalar glueball mass remains constant from $T=0$ to $T_c$ up to fluctuations of order 100 MeV. Notice that such a behavior was expected from calculations relying on effective low-energy lagrangians~\cite{lowe}. The decay width of the scalar glueball, however, has been shown to increase significantly because of thermal effects~\cite{ishi2,ishi3}. In the more recent work~\cite{meng}, the pseudoscalar and tensor glueball masses are also shown to be constant from $T=0$ to $T_c$, again up to an accuracy of 100 MeV.

How can those results be qualitatively explained within the relativistic flux tube model? First, let us consider that the dominant contribution of the thermal effects is to modify the string tension according to Eq.~(\ref{aeff}). Second, the influence of $T$ on the instanton-induced effects is neglected. Thus, in order to get a nearly constant scalar glueball mass, $m_g$ has to increase with $T$ in order to compensate the decrease of the string tension. The idea that gluons should gain a thermal mass has already been mentioned in Ref.~\cite{glumt} for example. Since no conclusion can be drawn about the form of $m_g(T)$, we consider it as a free parameter and compute it at various temperatures so that $\partial_TM_{0^{++}}=0$ between $T$ and $T_c$. The gluon masses numerically obtained are well fitted by the following form 
\begin{equation}\label{glum}
	m_g(T)=0.765\, \frac{T}{T_c}\left[1+0.486\left(\frac{T}{T_c}\right)^2\right]\ {\rm GeV}.
\end{equation}

We now turn to the thermal decay width of the scalar glueball. We gave in Sec.~\ref{decay} a simple way to estimate the glueball decay width in the case where the decay by pair creation is the dominant mechanism. Using the fitted form~(\ref{glum}) and the formula~(\ref{aeff}), it can be computed that $\sqrt{\left\langle r^2\right\rangle}$ significantly increases from 2.850~GeV$^{-1}$ at $T=0$ to $3.823$~GeV$^{-1}$ at $T=0.9\, T_c$. However, $a(T)$ quickly decreases, so that $a(T)\, \sqrt{\left\langle r^2\right\rangle}$ becomes negligible at high $T$, meaning that the decay by pair creation is no longer dominant. What we have instead is a progressive ``dissolution" of the glueball in the medium, expressed by the increase of $\sqrt{\left\langle r^2\right\rangle}$ with $T$; such a phenomenon should contribute to increase the decay width of the glueball. Moreover, since the constituent gluons become quite heavy near $T_c$, perturbative mechanisms such as those found in heavy quarkonia might come into play. It is known for example that, because of in-medium effects, heavy quarkonia gain a thermal decay width that roughly grows like $T$~\cite{Brambilla:2008cx}. Again, such a mechanism could increase the glueball decay width. We can thus conclude that an increase of the glueball decay width with $T$ is not excluded \textit{a priori} within our effective approach, but further theoretical investigations are needed to clarify and confirm that statement.    
 
To our knowledge, no widely accepted theoretical derivation of the thermal gluon mass between $T$ and $T_c$ can be found in the literature so far. We have shown that the form~(\ref{glum}) leads to the expected features for the $0^{++}$ state. An interesting point to mention is the following: Only a state-dependent thermal mass is able to keep all the glueball spectrum constant with respect to the temperature. It can be seen from the approximate mass formula~(\ref{2gma}) indeed that the whole glueball spectrum will have a constant mass provided that $m_g(T)$ is approximately given by
\begin{equation}\label{glum2}
	m_g(T)\approx\frac{M_0}{2\sqrt 2} \ \frac{T}{T_c}+O\left(\left(T/T_c\right)^3\right),
\end{equation} 
$M_0$ being given by Eq.~(\ref{2gmabis}), which moreover gives an \textit{a posteriori} justification of the form~(\ref{glum}). Further lattice studies concerning the evolution of the glueball masses versus the temperature in various channels could provide more information about the thermal gluon mass. If all the glueball masses appeared to be constant, then a state-dependent gluon mass like~(\ref{glum2}) would be needed. On the contrary, let us suppose that $m_g(T)$ is state-independent, but such that the scalar gluebal has a constant mass. Then, the pseudoscalar glueball will have a constant mass following our approach, since instanton-induced interactions are assumed to remain constant with the temperature, while the mass of the other glueball states will decrease with $T$. Indeed, the thermal mass is not large enough to balance the decrease of the string tension in that case. For example, using Eq.~(\ref{glum}), we compute that the $2^{++}$ mass could decrease from 2.529~GeV at $T=0$ to 2.326~GeV at $T=0.9\, T_c$. For the moment, lattice data do not allow to conclude that the tensor glueball mass significantly decreases; see~\cite{meng}. Computations with increased accuracy might be quite interesting in order to check whether that behavior is observed or not.
 
\section{Conclusions}\label{Conclu}

\begin{figure}[ht]
	\centering
		\includegraphics*[width=9cm]{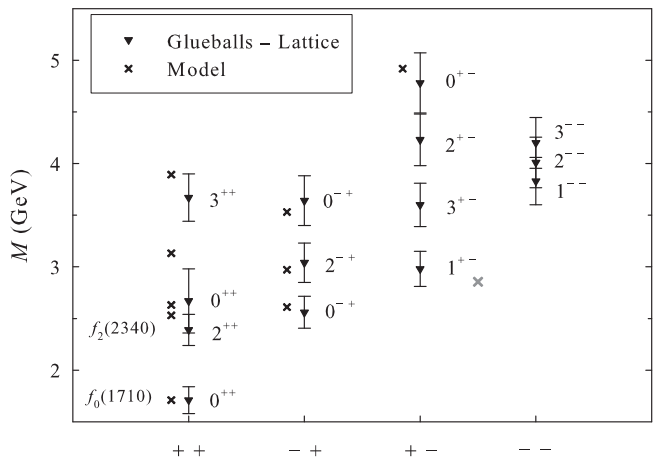}
	\caption{Summary plot of the data shown in Table~\ref{tab1} and in section~\ref{34g}. Lattice QCD data (triangles) from Refs.~\cite{latti0,latti1} are compared to the mass spectrum of the relativistic flux tube model (crosses). The lower bound~(\ref{mggg}) on the three-gluon glueball masses has been plotted in gray. }
	\label{figfin}
\end{figure}

The relativistic flux tube model has been firstly applied to compute masses and decay widths of glueballs seen as bound states of transverse constituent gluons. It allows not only to understand to qualitative features of the lattice spectrum, but leads also to a good quantitative agreement with these data. Figure~\ref{figfin} is a summary plot of the masses we obtain. Notice a peculiar result we have found: An open flux tube with massless particles attached at his ends has the same mass spectrum than a closed flux tube with an energy density two times smaller. There exists thus a kind of duality between open- and closed- flux tube models, that draws a bridge between apparently distinct approaches existing in the literature. 

Only a few calculations of glueball decay widths exist nowadays (see Refs.~\cite{abreu,cgglu} and references therein), and are mostly concerned with the scalar glueball. Our framework allows to estimate the decay width in various channels, leading in particular to the conclusion that the $f_2(2340)$ resonance could be a relevant candidate for the tensor glueball. Using recent KLOE data concerning the gluonic content of the $\eta'$, we are able to predict a mass formula for a pseudoscalar state that is pure glueball at 87\%. That state would have a mass around 2.9~GeV and a decay width around 300~MeV if we use the pure glueball mass computed in the present pure-glue model. But large sea-quark effects might play an important role in that sector to decrease the mass and maybe lead to identify the $\eta(1405)$ with the physical pseudoscalar glueball, as suggested in other recent works. The inclusion of sea-quark effects in effective models such as ours is actually a crucial issue for a better comparison to experimental results. 

We have also addressed the problem of the glueball-Pomeron identification. Our results suggest that the lowest-lying (even--$J$)$^{++}$ two-gluon glueballs should no be located on the leading Pomeron trajectory, but rather on a daughter trajectory. The leading trajectory could rather be associated to states with a nonzero quark-antiquark component; such a proposal demands further and careful studies of mixing effects at the nonperturbative level to be checked. 

Finally, we have proposed a way to include finite-temperature effects in our approach, reasonably assuming that thermal effects cause the string tension to decrease with $T$. Eventually the flux tube disappears at the critical temperature. Currently, lattice studies suggest that the scalar glueball mass remains constant between $T=0$ and $T_c$, while its decay width increases. The relativistic flux tube model seems to be able to reproduce qualitatively such a constant mass provided that the constituent gluons gain a thermal mass which increases with the temperature. An increase of the thermal decay width is not excluded \textit{a priori} by our approach, but further studies are needed to clarify the situation. Following that the thermal gluon mass is state-dependent or not, we predict the mass of the other glueball states to be respectively constant or significantly decreasing when approaching $T_c$; such a prediction could be checked in future lattice studies.  

\acknowledgments FB (F.R.S.-FNRS Postdoctoral Researcher) and CS (F.R.S.-FNRS Senior Research Associate) thank the F.R.S.-FNRS for financial support. VM thanks the IISN for financial support.

\end{document}